# A Spatio-Temporal Hybrid Quantum-Classical Graph Convolutional Neural Network Approach for Urban Taxi Destination Prediction

Xiuying Zhang , Qinsheng Zhu , and Xiaodong Xing

*Abstract*—We propose a Hybrid Spatio-Temporal Quantum Graph Convolutional Network (H-STQGCN) algorithm by combining the strengths of quantum computing and classical deep learning to predict the taxi destination within urban road networks. Our algorithm consists of two branches: spatial processing and time evolution. Regarding the spatial processing, the classical module encodes the local topological features of the road network based on the GCN method, and the quantum module is designed to map graph features onto parameterized quantum circuits through a differentiable pooling layer. The time evolution is solved by integrating multi-source contextual information and capturing dynamic trip dependencies on the classical TCN theory. Finally, our experimental results demonstrate that the proposed algorithm outperforms the current methods in terms of prediction accuracy and stability, validating the unique advantages of the quantum-enhanced mechanism in capturing high-dimensional spatial dependencies.

## I. INTRODUCTION

With technological advancements and the increasing ownership of vehicles, the pressure on public transportation has intensified, particularly in megacities with high population densities. The challenge is obvious: rapid urbanization and growing traffic congestion exacerbate the problem. In this context, accurate taxi destination prediction is vital. It improves vehicle dispatch system efficiency, minimizes parking search time, and contributes to the optimization of urban planning and infrastructure [1], [2], [3].

Destination prediction research has developed from early statistical methods, Markov chains, and traditional machine learning techniques that relied on shallow features such as speed and direction [4], [5], [6]. However, as trajectory data expands in scale and complexity, these shallow models fail to effectively learn the underlying complex patterns. This shift has led to the adoption of deep learning models, such as long short-term memory (LSTM) networks and temporal convolutional networks (TCN) [7], [8], [9]. These advancements have enhanced trajectory temporal modeling through stronger representation capabilities. Unfortunately, it struggles to effectively handle static road network topology and multi-level spatial dependencies, often compressing spatial features into a single vector [10], [11], [12].

To solve this problem, graph convolutional networks (GCN) and graph neural networks (GNN) have been widely adopted to aggregate neighborhood information via graph structures, thereby explicitly modeling spatial correlations in traffic prediction and path planning [8], [10], [13], [14]. Whereas, existing graph methods still encounter considerable limitations when dealing with high-dimensional, sparse, and dynamically evolving urban trajectory data, particularly regarding model depth, parameter efficiency, and dynamic spatiotemporal feature fusion [15], [16].

To overcome the limitations of classical computing in feature extraction, quantum computing offers a new approach in processing high-dimensional data. Based on the superposition and entanglement properties of qubits, quantum algorithms can map low-dimensional graph data into an exponentially dimensional Hilbert space, thereby capturing deep correlations and high-dimensional features that remain imperceptible to classical methods [4], [14], [19]. The proposal of hybrid quantum-classical neural network architectures [21] integrates parameterized quantum circuits (PQC) into key layers of classical neural networks. This approach overcomes the computational limitations of current noisy intermediate-scale quantum (NISQ) hardware. It also improves feature extraction and representation using quantum mechanisms [12], [17], [18]. Furthermore, quantum graph convolutional neural networks (QGCN) enable deeper feature modeling of graph-structured data through quantum state evolution, providing a practical path to resolve the high-dimensional feature mapping challenges faced by Graph convolutional network (GCN) [1], [4], [15], [19].

Motivated by these opportunities and challenges, we propose a H-STQGCN for urban taxi destination prediction. Our approach aims to leverage the high-dimensional mapping advantages of quantum computing to compensate for the

This research was funded by the Natural Science Foundation of Xinjiang Uygur Autonomous Region (No. 2024D01A17), and the Chengdu Key Research and Development Program (No. 2025-YF08-00109-GX). (Corresponding author: Qinsheng Zhu.)

Xiuying Zhang is with the School of Physics, University of Electronic Science and Technology of China, Chengdu 610054, China (e-mail: zxy02402@163.com).

Qinsheng Zhu is with the School of Physics, University of Electronic Science and Technology of China, Cheng Du, 610054, China and the Institute of Electronics and Information Industry Technology of Kash, Kash, 844000, China (e-mail: zhuqinsheng@uestc.edu.cn).

Xiaodong Xing is with the School of Quantum Information Future Technology, Henan University, Zhengzhou 450046, China, the Henan Key Laboratory of Quantum Materials and Quantum Energy, Henan University, Zhengzhou 450046, China, and the Institute of Quantum Materials and Physics, Henan Academy of Sciences, Zhengzhou, 450046, China (e-mail: xiaodong.xing@henu.edu.cn).

shortcomings of classical deep learning, constructing an end-to-end solution for vehicle destination prediction.

The main contributions of this paper are as follows. Firstly, we propose a hybrid spatial representation framework that fuses classical road network topology extraction with quantum high-dimensional feature mapping to enhance the representation capability of complex road network structures. Secondly, PQC-enhanced graph convolution and pooling layers are proposed to address the limitations of classical GCN in deep feature extraction and node representation. Finally, a collaborative fusion mechanism is designed between the extracted quantum spatial features and temporal dynamic modeling, achieving a unified characterization of non-local spatial dependencies and temporal evolution patterns.

The paper is organized as follows: Section II reviews the related work. Section III details the proposed algorithm. Section IV demonstrates the effectiveness of our approach through comparative experiments. Finally, Section V concludes the paper.

## II. RELATED WORK

This section introduces a concise overview of recent research advances in the field of ITS, focusing on destination prediction, spatio-temporal graph networks (ST-GCN) based traffic forecasting, and QGCN.

### A. Destination Prediction

Destination prediction refers to the accurate inference of a vehicle's final destination based on multi-source information, such as vehicle trajectories, historical travel records, weather conditions, and regional functionalities, thereby providing a decision-making basis for traffic dispatching [5], [23].

In the early stage of development, Mikluščák et al. [6] pioneered the introduction of neural networks into destination prediction, verifying the potential of deep learning applications. Yan et al. [24] proposed the concept of "semantic trajectory," which significantly enhanced the semantic understanding of prediction by annotating GPS data with regional functionalities. With the deepening of research methodologies, Xu et al. [5] proposed the DESTPRE algorithm, which utilizes trajectory clustering and feature matching to achieve prediction, incorporating "partial trajectory integrity" as a key consideration. Besse et al. [4] improved prediction robustness by modeling the trajectory probability distribution, whereas Jamil et al. [25] provided a new analytical perspective for prediction by starting from the demand side [26].

In recent years, deep learning and multimodal fusion have continuously promoted technical progress. Lv et al. [27] exploited convolutional neural networks (CNN) to capture spatial features, while Liao et al. [9] combined Bi-directional LSTM with attention mechanisms to mine fine-grained temporal information. Rossi et al. [28] introduced driver behavioural features, and Qian et al. [29] established a framework for long-distance trajectory correlation. At the application level, Xu et al. [31] embedded prediction algorithms into dispatching systems, realizing the joint optimization of "demand-destination." Research by Abideen et al. [32] and Guo et al. [1] effectively improved the applicability of algorithms in large-scale urban scenarios.

Overall, destination prediction has evolved from path matching to intelligent inference based on multi-source information, yielding significantly improved accuracy. Future works will be devoted to exploring directions such as multi-source fusion and algorithm lightweighting.

### B. Spatio-Temporal Graph Neural Networks in Traffic Forecasting

The core challenge in traffic prediction lies in capturing the complex dependencies between road network topology and time-varying characteristics [7], [23]. Spatio-temporal graph neural networks (ST-GNN) have become a mainstream paradigm by modeling the road network as a "node-edge" structure, effectively integrating the spatial extraction capabilities of GCN with the temporal modeling abilities of sequence algorithms, such as LSTM or GRU [33].

Early research focused on the fusion of multi-source information. For instance, ST-MGCN exploits distance and functional multi-view graphs to capture demand correlations, or embeds external contexts, such as weather and holidays, into the graph structure to enhance scene perception [12], [34]. In order to further improve the representation of dynamic correlations, attention mechanisms have been widely introduced. Zhang et al. [33] and Kumar et al. [13] achieved capturing of core road segment weights and long-short-term dependencies by employing graph attention layers and "spatial-temporal dual attention" structures, respectively. Moreover, addressing road network structural changes and multi-scale issues, algorithms such as Test-GCN and T-GCN have effectively enhanced prediction robustness by introducing topological similarity matrices and hierarchical aggregation strategies [11], [35], [36].

Recent research has focused on addressing the efficiency bottlenecks of complex algorithms. Works such as STC-PSSA and TC-GCN introduce probabilistic sparse self-attention and multi-hop mechanisms, significantly reducing computational costs while selecting key nodes and modeling indirect spatial associations [7], [8], [10].

Although ST-GNN have effectively improved prediction accuracy, computational redundancy remains when dealing with large-scale dynamic road networks. Future research should be devoted to exploring lightweight deployment and online updating of dynamic graphs.

### C. Quantum Graph Convolutional Neural Networks

To address the computational bottlenecks of high-dimensional traffic road networks, quantum computing provides an efficient parallel computing paradigm for graph neural networks (GNN) by virtue of superposition and entanglement characteristics [15], [17]. Given the limitations of current NISQ devices, the hybrid quantum-classical architecture has become a mainstream solution. Chen et al. [16] and Fan et al. [20] effectively reduced hardware load and noise interference by separating quantum feature extraction

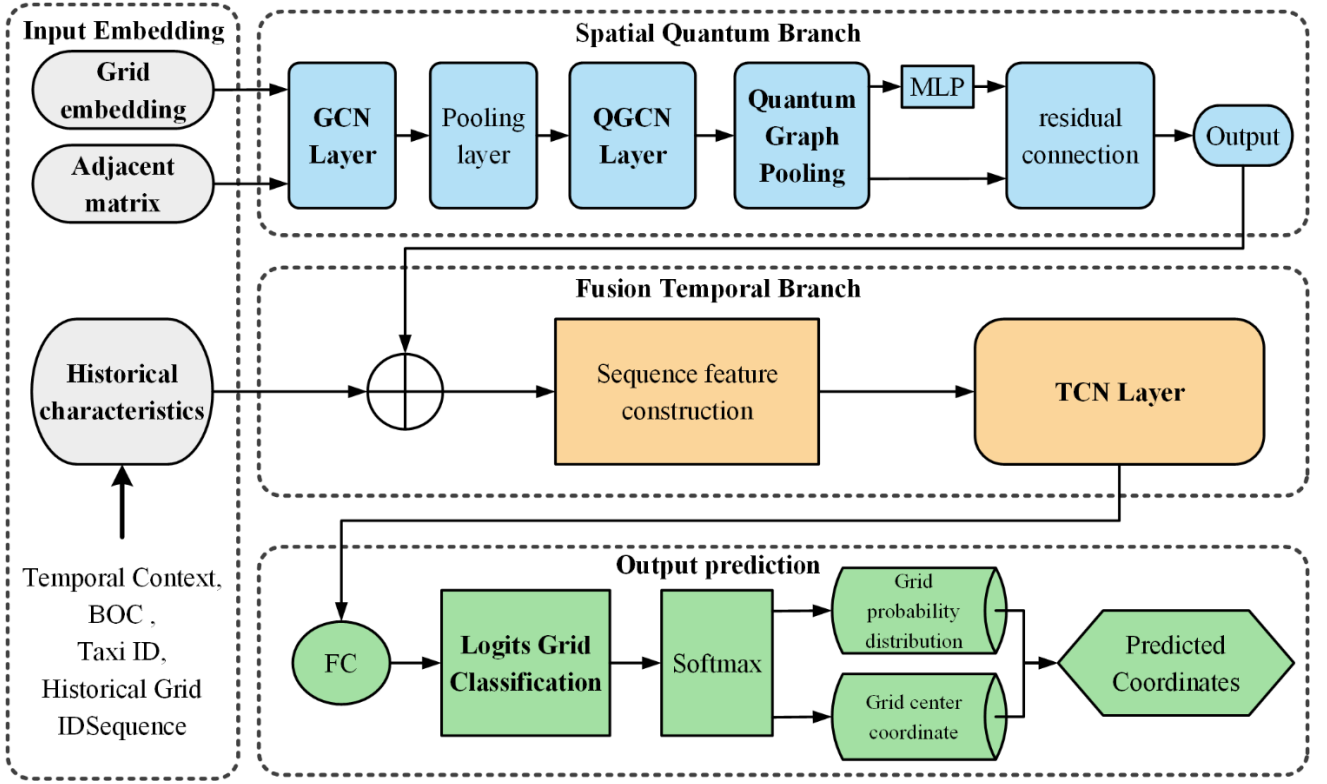

Fig. 1. Architecture and Framework of the H-STQGCN Algorithm. The network consists of an Input Embedding module, a Spatial Quantum Branch, a Fusion Temporal Branch, and an Output Prediction module. Part a: Input Embedding for initializing graph inputs and historical characteristics. Part b: Spatial Quantum Branch for encoding city graphs and extracting deep spatial features via GCN and QGCN. Part c: Fusion Temporal Branch for capturing dynamic sequence patterns using TCN. Part d: Output Prediction for generating grid probabilities and final coordinates.

from classical parameter updating.

In the context of specific traffic modeling, researchers have attempted to integrate quantum mechanisms into spatio-temporal prediction frameworks. Qu et al. [18] pioneered the spatio-temporal quantum graph convolutional network, exploiting quantum state superposition to characterize the multi-state uncertainty of traffic congestion. Subsequently, addressing complex spatial dependencies, Zheng et al. [17] and Bai et al. [37] introduced quantum entanglement mechanisms, significantly enhancing information propagation and correlation modeling capabilities among nodes. In addition, explorations based on spectral graph theory are also deepening; for instance, the QGCN mathematical framework established by Zheng et al. [19] and the spectral domain adaptive feature extraction method proposed by Ye et al. [22] have provided new insights for dynamic road network prediction.

Overall, quantum computing provides a new paradigm for traffic prediction to overcome computational limits. However, future research should be devoted to deepening the study of collaborative optimization and hardware adaptation.

## III. METHODLOGY

The objective of predicting taxi destinations is to accurately determine the endpoint coordinates of a trip based on the vehicle's historical trajectory, contextual information, and the complex intrinsic structure of the urban geodata environment. This problem can be formalised as a sequence prediction and sequence-to-point prediction task, which is solved by a joint classification-regression strategy. The algorithm commences with the historical trajectory sequence $S_k = \{(G_{k,t-H}, C_{k,t-H}), \ldots, (G_{k,t-1}, C_{k,t-1})\}$ of a taxi and the current contextual features $F_{Contxt}$. It then predicts the discrete grid ID of the destination in the subsequent time step t and subsequently regresses the final continuous geographic coordinate destination $\hat{Y}_{k,t} = (\widehat{lon}_{k,t}, \widehat{lat}_{k,t})$ based on this classification result. In this context, $G$ denotes the city grid ID sequence and $C$ denotes the semantic features of the corresponding grid.

The proposed H-STQGCN algorithm, whose architecture is shown in Fig 1, comprises three core modules:

a) The spatial quantum branch first uses GCN to encode large-scale city graphs and then coarsens these graphs via a differentiable graph pooling layer to transform them into the quantum domain. Next, QGCN and quantum graph pooling extract quantum-enhanced spatial vectors $V_{Global}$ with global topological structure.

b) The temporal fusion branch integrates the global spatial vector $V_{Global}$ into each grid feature of the trajectory sequence. Combined with contextual features such as taxi ID and temporal information, it captures dynamic sequence patterns via TCN to generate the temporal feature vector $V_{Seq}$.

c) Finally, the prediction module maps $V_{Seq}$ to the logit space of grid IDs. By incorporating the geographic coordinates of the grid center and performing a weighted

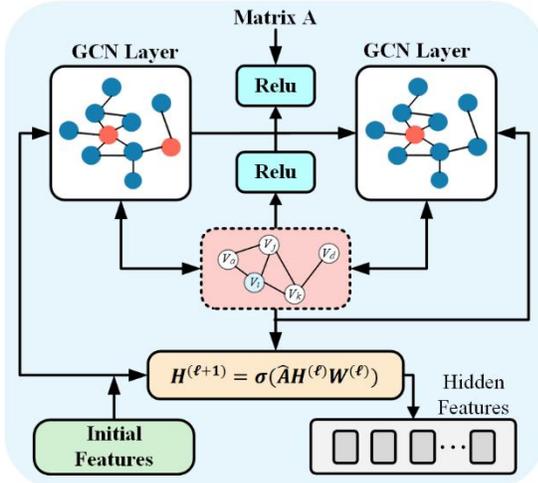

Fig. 2. Convolutional Neural Network. The module transforms initial grid embeddings into high-dimensional hidden features using the normalized adjacency matrix $\hat{A}$ and residual connections. It employs the layer-wise propagation rule $H^{(\ell+1)}$ to capture complex local topological dependencies among urban nodes.

summation, it delivers the final predicted coordinates.

### A. Trajectory Feature Engineering and Spatial Topology Modeling

*a) Feature engineering for trajectories*

The core concept of feature engineering is to convert raw, continuous GPS trajectory points into discrete, semantically rich structured inputs. Spatially, the entire urban area is discretized into uniformly sized grid cells, with all historical trajectory points (start and end points) assigned to corresponding grid ID sequences. To quantitatively characterize the urban functional attributes of each grid (e.g., commercial, residential, transportation hubs), a bag-of-categories (BOC) vector is introduced based on the distribution of points of interest (POI). This vector is derived by statistically calculating and normalizing the frequency of occurrence $n_i(g)$ for $K$ categories of POIs within the grid $g$:

$$v_{BOC}(g) = [n_1(g), n_2(g), \ldots, n_K(g)] / \sum_{i=1}^{K} n_i(g) \quad (1)$$

At the same time, each discrete grid ID is mapped to a low-dimensional dense vector via an independent embedding layer, allowing the neural network to learn spatial correlations during training. Along the temporal dimension, contextual information such as the hour, day of the week, and type of day of the week on which the trajectory occurred is also processed via corresponding embedding layers. Ultimately, the concatenation of the spatial grid embedding, the BOC vector, and the temporal context embedding at each historical trajectory point together form the comprehensive feature representation of the trajectory and provide subsequent algorithms with holistic spatiotemporal dynamic information.

*b) Spatial topological modeling*

To effectively capture the complex spatial structure of cities, the algorithm introduces an adjacency matrix based on geographic proximity to characterize the static spatial relationships between urban grids. This diagram treats each grid as a node and uses the Haversine formula to accurately calculate the geographic distance between any two grid centers, $i$ and $j$:

$$a = \sin\left(\frac{lat_i - lat_j}{2}\right)^2$$

$$+ \cos(lat_i)\cos(lat_j)\sin\left(\frac{lon_i - lon_j}{2}\right)^2 \quad (2)$$

$$d_{Haversine}(i,j) = 2r \arcsin(\sqrt{a}) \quad (3)$$

If this distance falls below a preset threshold, an edge is created between the two nodes to simulate strong spatial dependencies within the local region. To ensure the stability and numerical balance of the GCN during the aggregation of neighborhood information, the generated adjacency matrix $A$ undergoes symmetric normalization processing:

$$\hat{A} = D^{-\frac{1}{2}}(A + I)D^{-\frac{1}{2}} \quad (4)$$

$$D_{ii} = \sum_j (A_{ij} + I_{ij}) \quad (5)$$

Ultimately, this normalized adjacency matrix $\hat{A}$, which embodies the static spatial structure of the city, and the initial feature representation of the grid are fed together into the spatial quantum branch of the algorithm. This serves to extract deep spatial structural dependencies between the grid cells.

### B. Spatial Quantum Branch

The Spatial Quantum Branch is responsible for extracting high-order spatial features from the city grid's topological structure. Its objective is to learn a vector that is independent of input sequence length and capable of encoding global spatial-topological information.

GCN form the cornerstone of spatial feature extraction, with the core objective of capturing complex local spatial dependencies among nodes in urban grids, as illustrated in Fig 2. During the initial stage of the algorithm, each grid cell is represented by a learnable embedding vector. The vectors of all nodes collectively form the initial feature matrix $X^{(0)} \in \mathbb{R}^{N \times D_{in}}$, where $N$ denotes the total number of grid nodes and $D_{in}$ represents the initial embedding dimension. This feature matrix is subsequently fed into a deep network composed of stacked GCN layers. GCN propagates node features through adjacency matrices according to the layer-wise computation:

$$H^{(\ell+1)} = \sigma(\hat{A}H^{(\ell)}W^{(\ell)}) \quad (6)$$

where $H^{(0)} = X^{(0)}$, $\hat{A} = \tilde{D}^{-1/2}\tilde{A}\tilde{D}^{-1/2}$ denotes the normalized adjacency matrix, $\tilde{A} = A + I_N$ represents the adjacency matrix with self-loops included, $W^{(\ell)}$ is the trainable weight matrix, and $\sigma(\cdot)$ is the nonlinear activation function. Therefore, to improve the training stability and avoid gradient vanishing issue in deeper GCN, we add residual connection as follows $X_{out} = ReLU(\hat{A}X_{in}W + Downsample(X_{in}))$ to make sure nodes keep their information in feature propagation process. Finally, after multiple layers of graph convolutional processing, GCN could learn the local topological structure among nodes and output high-dimensional hidden feature matrix $X_{GCN} \in \mathbb{R}^{N \times D_{hidden}}$ to provide strong spatial representation support for following spatio-temporal feature fusion or quantum feature extraction process.

However, the quantity of qubits in current quantum

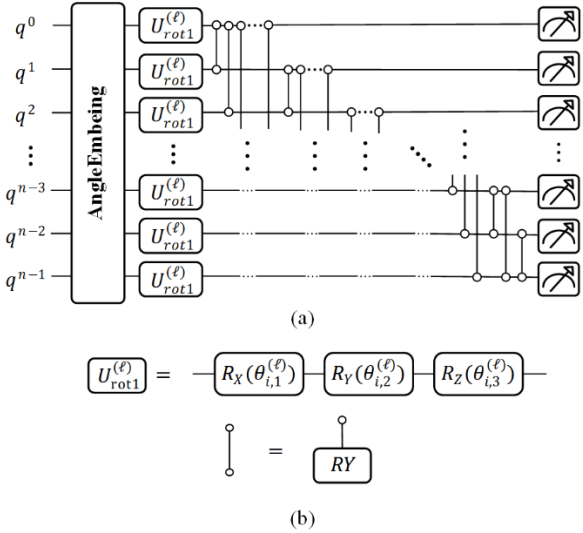

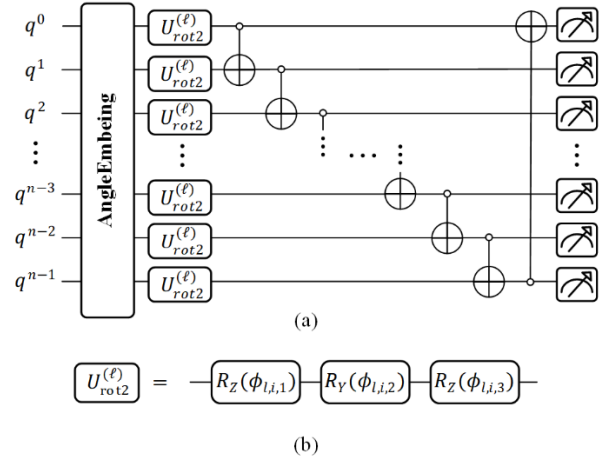

Fig.3. Quantum Circuit for Quantum Graph Convolution Layer. (a) Parametric quantum circuit diagram for quantum graph convolutional networks. (b) Subcircuit diagram of a parameterized quantum circuit.

Fig. 4. Quantum Graph Pooling Layer Quantum Circuit. (a) Quantum Circuit Diagram of the Quantum Pooling Layer. (b) Subcircuit diagram of a parameterized quantum circuit

computers is limited, which hinders the direct processing of large graph data. Therefore, we apply a differentiable graph pooling layer after GCN to hierarchically compress the graph in an end-to-end way while feature representation is still preserved. The differentiable graph pooling layer learns the cluster assignment matrix $S \in R^{N \times N_{qubits}}$ via an independent GCN module, where $N_{qubits}$ denotes the target number of nodes (i.e., qubits in a quantum circuit). The assignment matrix is calculated as $S = softmax(GCN_{assign}(\hat{A}, X_{GCN}))$, where $S_{ij}$ represents the probability of node $i$ being assigned to cluster $j$. Subsequently, this assignment matrix is used to pool node features and graph structure:

$$X_{Pooled} = S^T X_{GCN} \in R^{N_{qubits} \times D_{hidden}} \quad (7)$$

$$A_{Pooled} = S^T \hat{A} S \in R^{N_{qubits} \times N_{qubits}} \quad (8)$$

This process compresses the original $N$ nodes into $N_{qubits}$ cluster nodes while generating a new feature matrix $X_{Pooled}$ and a coarsened adjacency matrix $A_{Pooled}$, where $A_{Pooled}(i,j)$ denotes the connection strength between clusters $i$ and clusters $j$.

QGCN layer is an important part of spatial quantum branch. QGCN aims to use the superposition and entanglement advantages of quantum computing to realize high-dimensional and nonlinear feature mapping of graph structured data in the quantum state space. Unlike traditional graph convolution methods which only spread information based on adjacency matrix, spatial quantum branch uses the parameterized gate operation in quantum circuits to correlate the quantum state between nodes.

This approach enables the algorithm to capture more profound topological dependencies and nonlinear relationships, thereby enhancing the capacity for spatial feature representation. As shown in Fig 3, the node feature matrix $X_{Pooled} \in R^{N_{qubits} \times D_{hidden}}$ is first mapped to the qubit space via the differentiable pooling layer, where each node corresponds to a qubit. Node features are encoded as quantum states using angle coding:

$$|\psi_{in}\rangle = \bigotimes_{i=1}^{N_{qubits}} R_Y(x_i) R_Z(x_i) |0\rangle_i \quad (9)$$

where $x_i$ denotes the node feature, and $R_Y$ and $R_Z$ represent rotation operations around the Y-axis and Z-axis, respectively.

This encoding ensures that input features are embedded as phase angles within the probability amplitudes of quantum states, thereby achieving a lossless mapping from classical vectors to quantum states. Subsequently, a parameterized quantum circuit is employed to propagate features via quantum convolution. This quantum convolution operation can be formally expressed as: $|\psi_{out}\rangle = U_{QGCN}(\Theta_{rot}, \Theta_{ent}, A_{Pooled}) |\psi_{in}\rangle$, where $U_{QGCN}$ denotes the overall propagation operator, which is composed of alternating rotation and entanglement layers:

$$U_{QGCN} = \prod_{\ell=1}^{L} [U_{rot}^{(\ell)}(\Theta_{rot}^{(\ell)}) \cdot U_{ent}^{(\ell)}(\Theta_{ent}^{(\ell)}, A_{Pooled})] \quad (10)$$

The rotation layer applies single-qubit gate transformations to simulate phase evolution of the node's intrinsic features:

$$U_{rot}^{(\ell)} = \prod_{i=1}^{N_{qubits}} R_X(\theta_{i,1}^{(\ell)}) R_Y(\theta_{i,2}^{(\ell)}) R_Z(\theta_{i,3}^{(\ell)}) \quad (11)$$

where $\Theta_{rot}^{(\ell)}$ denotes the learnable rotation parameters of layer $\ell$. The entanglement layer establishes quantum correlations between qubits via controlled-rotation gates, with coupling strengths dynamically modulated by the pooled adjacency matrix $A_{Pooled}$:

$$U_{ent}^{(\ell)} = \prod_{(i,j) \in E} CRY(\theta_{ij}^{(\ell)} \cdot A_{Pooled}(i,j)) \quad (12)$$

where CRY denotes the controlled-Y rotation gate. This graph-structured modulation mechanism ensures that entanglement strength between nodes corresponds to their topological connectivity, thereby preserving spatial consistency during the evolution of quantum features.

After the quantum propagation concludes, the output state $|\psi_{out}\rangle$ is measured using the Pauli Z operator to extract observables: $z_i = \langle \psi_{out} | Z_i | \psi_{out}\rangle, i = 1,2,\ldots,N_{qubits}$. This procedure yields the quantum feature response vector:

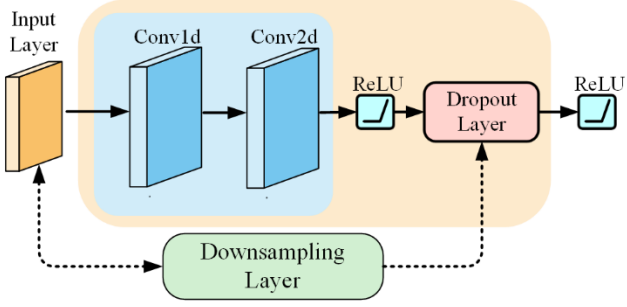

Fig. 5. Temporal Convolutional Network. The module uses cascaded conv layers with ReLU and Dropout for temporal feature extraction. A residual connection with a Downsampling Layer is incorporated to ensure dimension alignment.

$Z_{quantum} = [z_1, z_2, \ldots, z_{N_{\text{qubits}}}]^\top$. These measurement results are subjected to linear transformation and nonlinear activation mapping to generate the quantum convolutional output feature matrix:

$$X_{QGCN} = \text{ReLU}(W_{out} Z_{quantum} + X_{Pooled}) \quad (13)$$

where the residual term ensures information fusion between quantum and classical features and maintains gradient flow stability. From the perspective of quantum physics, the propagation process of quantum graph convolution is equivalent to a unitary evolution governed by the graph Hamiltonian $H_{\text{graph}}$ in quantum state space:

$$|\psi(t+1)\rangle = e^{-iH_{\text{graph}}\Delta t}|\psi(t)\rangle \quad (14)$$

The coupling terms of $H_{\text{graph}}$ are determined by the adjacency matrix $A_{\text{Pooled}}$, representing energy coupling relationships between nodes and enabling modeling of both local topology and global spatial dependencies.

The quantum pooling layer (Fig 4) further aggregates the features produced by QGCN into a global vector $V_{\text{RawGlobal}}$, thereby achieving feature compression and aggregation within the quantum state space. First, node-level features are averaged to obtain the input vector $x_{\text{mean}}$, which is then projected onto the qubit space via a linear mapping layer to form the input state:

$$|\psi_{\text{out}}\rangle = U_{\text{Embed}}(x_{\text{mean}})|0\rangle^{\otimes N_{\text{qubits}}} \quad (15)$$

$$U_{\text{Embed}} = \prod_{i=1}^{N_{\text{qubits}}} R_Y(x_i) \quad (16)$$

Subsequently, the pooling operation is implemented by the parameterized operator $U_{QPool}(\Phi_{pool})$, which performs quantum feature mixing through a multi-layer strongly entangled structure, expressed as:

$$U_{QPool} = \prod_{l=1}^{L} \left( \prod_{i=1}^{N_{\text{qubits}}} R_Z(\phi_{l,i,1}) R_Y(\phi_{l,i,2}) R_Z(\phi_{l,i,3}) \right)$$
$$\left( \prod_{i=1}^{N_{\text{qubits}}} \text{CNOT}(i, (i+1) \bmod N_{\text{qubits}}) \right)$$

$$(17)$$

where $\Phi_{pool}$ represents a set of learnable parameters. After quantum evolution, the output state $|\psi_{\text{pool}}\rangle = U_{QPool}(\Phi_{pool})|\psi_{\text{out}}\rangle$ is obtained. A Pauli $Z$ measurement is then applied to each qubit, yielding the pooled features $x_i^{(pool)} = \langle \psi_{\text{pool}}|Z_i|\psi_{\text{pool}}\rangle$, which constitute the quantum pooled feature vector:

$$X_{QPool} = [x_1^{(pool)}, \ldots, x_{N_{\text{qubits}}}^{(pool)}]^\top \quad (18)$$

This feature is input into a Multi-Layer Perceptron for nonlinear mapping, and through residual connections, it is added to the original global vector $V_{\text{RawGlobal}}$ to generate the final global vector $V_{\text{Global}}$. This process achieves hierarchical compression of quantum features and condensation of graph-level information while ensuring the stability and depth of feature learning.

*C. Temporal Fusion Branch*

The Temporal Fusion Branch is designed to capture the dynamic evolution and sequential dependencies of taxi trajectories while integrating multi-source contextual information to improve prediction performance.

The global vector $V_{\text{Global}}$ extracted from the Spatial Quantum Branch is first utilized to augment the original grid embedding $E_{\text{Grid}}$, thereby generating a graph-aware grid embedding $E'_{\text{Grid}} = \text{Linear}([E_{\text{Grid}}, V_{\text{Global}}])$. This enables each grid node to integrate higher-order information derived from the overall spatial structure while retaining its local features. Subsequently, the enhanced grid embeddings are concatenated with semantic and contextual features to form a sequence input matrix:

$$F_{\text{Seq}} = [E'_{\text{Grid}}, E_{\text{BOC}}, E_{\text{TaxiID}}, E_{\text{Hour}}, E_{\text{Weekday}}, E_{\text{DayType}}]$$
$$(19)$$

The input is designed to simultaneously encompass trajectory space, semantic constraints, and temporal context, thereby providing a comprehensive representation for sequence modeling. The sequence matrix $F_{\text{Seq}} \in \mathbb{R}^{D_{\text{feature}} \times L_{\text{sequence}}}$ is then fed into a TCN, as shown in Fig 5, which is composed of multi-layer residual convolutional blocks and dilated convolutions. Each layer produces the following output:

$$H^{(l)} = \sigma(\text{Conv1D}^{(l,2)}_{\text{dilated}}(\sigma(\text{Conv1D}^{(l,1)}_{\text{dilated}}(H^{(l-1)}))) + Dropout(\cdot)) + R^{(l)}$$

$$(20)$$

where the residual connections are defined as:

$$R^{(l)} = \{ \begin{array}{l} H^{(l-1)} \\ W_{\text{down}} H^{(l-1)} \end{array} \quad (21)$$

This design ensures gradient stability in deep networks, while dilated convolutions regulate kernel sampling intervals through the dilation factor, allowing an exponential increase in receptive field size across layers. Ultimately, the final time-step vector of the TCN output matrix $V_{TCN}$ is extracted to serve as the trajectory sequence's final feature representation $V_{Seq} = V_{TCN}[:, -1]$. This representation integrates historical trajectory information, enriched spatial–topological features, and temporal as well as individual contextual information, thereby providing a high-dimensional, spatiotemporally consistent dynamic representation for destination prediction—facilitating efficient sequence modeling and robust forecasting performance.

## D. Output Prediction Module

The output prediction module transforms the spatiotemporally fused features extracted by the dual-branch architecture into the final destination predictions. The sequence feature representation $V_{\text{Seq}}$ is first processed by a fully connected layer to obtain logit distributions $Z = \text{FC}(V_{\text{Seq}})$ for all grid cells, which are then converted into predicted probabilities $P_p = \text{Softmax}(Z)$ for the target grid cells. Here, $P_{p,i}$ denotes the probability that the $i$th grid cell is predicted to be the taxi's next destination. The algorithm maps discrete probabilities onto continuous coordinates using a pre-defined grid center lookup table $C = \{C_i \mid i = 1,2,\ldots,N_{\text{grids}}\}$, yielding the final predicted coordinates:

$$\hat{Y} = \sum_{i=1}^{N_{\text{grids}}} P_{p,i} \cdot C_i \tag{22}$$

This approach achieves a unified modeling framework that bridges classification and regression tasks. During training, end-to-end optimization is performed to minimize the cross-entropy loss between the logits and the true grid labels, thereby ensuring consistency between spatial localization accuracy and the predicted probability distribution.

## IV. EXPERIMENT

### A. Datasets

To further analyze the generalization ability and robustness of the proposed algorithm in different urban transportation environments, we choose three typical real-world taxi trajectory datasets from New York City, San Francisco and Porto.

*a) New York City Dataset:* Downloaded from the 2013 NYC Yellow Taxi Open Data repository. The data is organized by months. We choose the data of first three months, and due to the geographic limitation, we only keep the trips originated and terminated within the core area of Manhattan. Finally, we randomly choose 600 active drivers, which result in 646,909 trip records in total. Each trip contains pickup and dropoff latitude/longitude, timestamp and taxi ID. It can be regarded as a typical OD data structure.

*b) San Francisco Dataset:* Downloaded in 2008. It contains 464,019 complete GPS trajectories of 536 taxis at 10 seconds per trace.

*c) Porto, Portugal Dataset:* Downloaded from ECML/PKDD 2015 Taxi Trajectory Prediction Challenge. It includes 1.7 million trip records collected from 442 taxis from July 2013 to June 2014. The GPS points are recorded every 15 seconds. Each trip contains origins and destinations as well as intermediate points, and metadata including taxi ID, call type, trip start time and date type (weekday or holiday).

To ensure the comparability and learnability of trajectory data from multiple cities, the dataset in this article is further processed according to the following steps:

*1) Geographic filtering:* Using OpenStreetMap boundaries, central urban traffic areas (e.g., Manhattan, downtown San Francisco, central Porto) are selected to focus on key analysis scenarios.

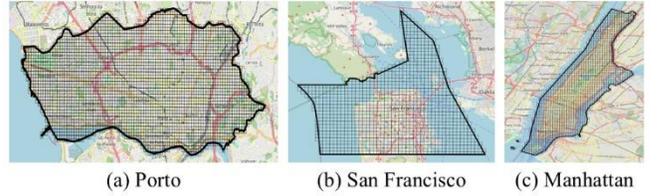

(a) Porto  (b) San Francisco  (c) Manhattan

Fig. 6. Grid Division Maps for Each City. Spatial discretization of central areas for (a) Porto, (b) San Francisco, and (c) Manhattan, using differentiated grid sizes adapted to local trajectory density.

TABLE I
TRAJECTORIES, GRIDS, AND POI COUNTS FOR THREE CITIES

| City | Number of Trajectories | Number of Grids | Number of POIs |
|---|---|---|---|
| Porto | 155877 | 3322 | 64931 |
| San Francisco | 73450 | 2023 | 197796 |
| Manhattan | 111327 | 1998 | 94535 |

*2) Building the trip sequence:* Based on the vehicle's historical trips, consecutive trips with boarding times within three hours of each other are grouped into a single complete trajectory segment. With a set input sequence length $L = 4$, the destination of the fifth segment is predicted based on the four previous historical trips, thereby unifying the modeling framework.

*3) Differentiated grid division:* Grid sizes are adjusted to the geographical size and trajectory density of each city – New York 218 m × 218 m, San Francisco 570 m × 570 m, Porto 115 m × 115 m. This achieves spatial discretization while aligning scaling, paving the way for uniform feature representation. The results of the grid partitioning for each city are shown in Fig 6.

*4) Spatial semantic enhancement:* Extract POI-related BOC semantic vectors for each grid. Quantify functional attributes by analyzing the distribution of primary POI categories within each grid, allowing the algorithm to simultaneously capture spatiotemporal patterns in trajectories and spatial functional differences between urban areas.

Following the procedures outlined above, the final statistical results for each city are presented in TABLE I, providing a reliable basis for subsequent algorithm training and performance evaluation.

### B. Implementation Details and Setup

This study realizes and trains a hybrid quantum–classical network on the Python platform through the assistance of PyTorch and PennyLane. The experiments are conducted on a 64-bit system with Intel i9-13900HX (2.20 GHz) and 16 GB of memory.

The whole algorithm is formed by spatial quantum branch and temporal fusion branch. For spatial branch, there are three layers of classical GCN and two layers of 8-qubit hybrid QGCN, which are responsible for extracting and mapping high-dimensional spatial features. For temporal branch, nine historical trajectory sequences are fed into the algorithm, and multi-scale temporal dependencies are learned by three TCN layers with dilation rates (1, 2, 4). The input features contain Taxi ID, temporal context and POI representation. The distance threshold of constructing adjacency matrix is tuned to be $\tau = 1.5$ kilometers.

TABLE II
EDS (KM) RESULTS FOR THREE CITIES

| Algorithm | Porto EDS | San Francisco EDS | Manhattan EDS |
|---|---|---|---|
| ARIMA | 2.3885 | 2.5356 | 2.8684 |
| NN | 2.3829 | 2.4186 | 2.7940 |
| MMLP-SEQ[28] | 2.2922 | 2.3455 | 3.6315 |
| LSTM[28] | 2.2700 | 2.4156 | 3.2498 |
| LSTM(BOC)[28] | 2.1813 | 2.2969 | 2.7178 |
| QLSTM[38] | 2.1140 | 2.2213 | 2.8629 |
| ST-GCN | 2.0600 | 2.1414 | 2.9163 |
| **H-STQGCN** | **2.0423** | **2.1573** | **2.6282** |

TABLE III
RMSE (KM) RESULTS FOR THREE CITIES

| Algorithm | Porto RMSE | San Francisco RMSE | Manhattan RMSE |
|---|---|---|---|
| ARIMA | 2.7815 | 3.0240 | 3.2647 |
| NN | 2.8120 | 2.9362 | 3.8976 |
| MMLP-SEQ[28] | 2.6945 | 2.8074 | 4.0646 |
| LSTM[28] | 2.6991 | 2.9933 | 3.8777 |
| LSTM(BOC)[28] | 2.6178 | 2.6990 | 5.8816 |
| QLSTM[38] | 2.4910 | 2.6488 | 3.2266 |
| ST-GCN | 2.3902 | 2.5504 | 3.6777 |
| **H-STQGCN** | **2.3134** | **2.5915** | **3.1608** |

In the training procedure, the Adam optimizer is adopted with learning rate of $1e-5$, batch size of 64 and cross-entropy loss function. The source data set is divided into 65% training subset, 15% validation subset and 20% testing subset by Taxi ID. The algorithm weights are updated and saved at the instant when the validation performance is optimal, which generalization capability.

*C. Evaluation Metrics*

This study utilizes two metrics—Euclidean Distance Error (EDS) [28] and Root Mean Square Error (RMSE)—to quantitatively assess and compare the performance of algorithms in multi-city taxi destination prediction tasks. EDS is adopted as the primary metric for experimental evaluation.

*a)* EDS is employed as the core metric for evaluating the geospatial prediction accuracy of the algorithm. It is calculated using the Haversine distance between the predicted coordinates $(\widehat{lon}, \widehat{lat})$ and the actual coordinates $(lon, lat)$, which measures the great-circle distance between two points in kilometers (km). EDS is defined as follows:

$$EDS = \frac{1}{N}\sum_{i=1}^{N} HaversineDistance(\hat{Y}_i, Y_i) \quad (23)$$

where $N$ denotes the total number of samples, $Y_i$ represents the true coordinates of the $i$ th sample, and $\hat{Y}_i$ indicates the algorithm's predicted coordinates. A lower EDS value corresponds to smaller geospatial prediction errors and higher prediction accuracy.

*b)* RMSE is primarily used to evaluate the algorithm's overall prediction stability and numerical error performance. It quantifies the average magnitude of prediction errors by computing the square root of the mean squared differences between predicted and actual results. RMSE is defined as follows:

$$RMSE = \sqrt{\frac{1}{N}\sum_{i=1}^{N}(\hat{Y}_i - Y_i)^2} \quad (24)$$

where $\hat{Y}_i$ and $Y_i$ represent predicted and actual values respectively. A smaller RMSE value suggests that the algorithm's predicted geographic coordinates are closer to the true values, thereby implying lower prediction errors.

*D. Experimental Results and Analysis*

To investigate the effectiveness and generalization capability of the proposed H-STQGCN in multi-city traffic forecasting, we conduct extensive experiments on three real-world taxi trajectory datasets from Porto, San Francisco, and Manhattan, New York with different road network structures and traffic conditions. We select several baseline algorithms for comparison, including traditional time-series algorithm (ARIMA), basic neural networks (NN, MMLP-SEQ), standard LSTM and its variant with BOC semantic features, as well as conventional ST-GCN.

It is worthwhile to note that the forward-looking quantum long short-term memory (QLSTM) network is adopted as a benchmark algorithm to explore the application of quantum computing in spatio-temporal forecasting. The EDS and RMSE evaluation metrics of all algorithms on three city datasets are presented in TABLES II and III respectively.

*a)* EDS experimental results

TABLE II presents the experimental results, indicating that the traditional time series algorithm ARIMA exhibits relatively high prediction errors, with EDS values of 2.3885 km in Porto, 2.5356 km in San Francisco, and 2.8684 km in New York. It struggles to capture the complex, nonlinear characteristics inherent in traffic flow. The basic neural network (NN) performs slightly better than ARIMA but remains limited in its ability to algorithm spatio-temporal dependencies. Simple fully connected architectures cannot effectively capture complex spatio-temporal correlations.

Deep sequence algorithms markedly enhance predictive performance; for instance, the LSTM achieves EDS values of 2.2700 km in Porto and 2.4156 km in San Francisco, validating the effectiveness of recurrent neural networks in modeling temporal dependencies. The LSTM incorporating BOC semantic features (LSTM(BOC)) exhibits superior performance. For the New York dataset, the EDS decreases from 3.2498 km to 2.7178 km—a 16.4% reduction—confirming the substantial contribution of urban functional zone semantics to improving prediction accuracy.

The QLSTM algorithm, representing the forefront of quantum computing, achieves outstanding performance by reducing [38] the EDS for the Porto dataset to 2.1140 km—an improvement of approximately 6.9% over LSTM. It further attains an EDS of 2.2213 km on the San Francisco dataset, representing an improvement of approximately 8.0% over LSTM. These findings validate the potential of quantum architectures for processing complex time series, highlight the unique advantages of quantum computing, and offer boththeoretical and practical guidance for the design of quantum–classical hybrid architectures.

TABLE IV
MELTING EXPERIMENT RESULTS FOR THREE CITIES

| | Algorithm | Porto EDS | Porto RMSE | San Francisco EDS | San Francisco RMSE | Manhattan EDS | Manhattan RMSE |
|---|---|---|---|---|---|---|---|
| A | GCN+TCN(without BOC) | 2.0696 | 2.4014 | 2.2877 | 2.6746 | 3.0154 | 3.5498 |
| B | GCN+TCN(with BOC) | 2.0600 | 2.3902 | 2.1414 | 2.5504 | 2.9163 | 3.6777 |
| C | GCN+QGCN+TCN(without BOC) | 2.0438 | 2.3646 | 2.1649 | 2.5746 | 2.6470 | 3.1117 |
| **D** | **GCN+QGCN+TCN(with BOC)** | **2.0423** | **2.3134** | **2.1573** | **2.5915** | **2.6282** | **3.1608** |

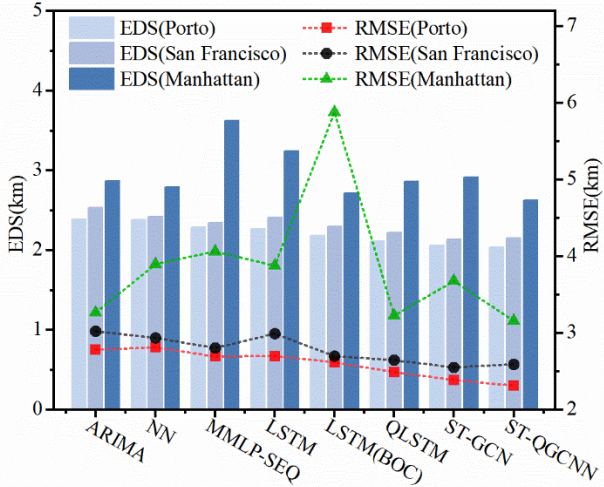

Fig. 7. EDS and RMSE Discounted Histogram. Comparative analysis of EDS (bars) and RMSE (lines) across three cities, highlighting the superior performance of the H-STQGCN algorithm.

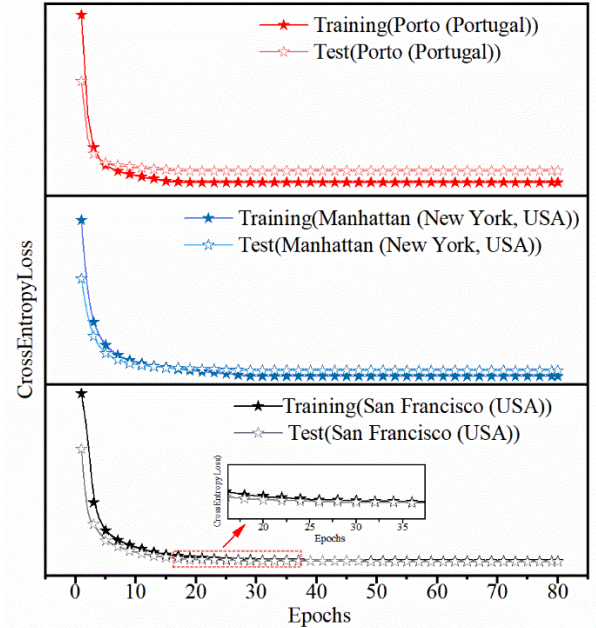

Fig. 8. Loss Images for Training and Validation Sets. This figure presents the loss images of the validation set and training set under three datasets.

The graph neural network ST-GCN exhibits superior performance, reducing EDS by 5.6% and 6.8% on the Porto and San Francisco datasets, respectively, relative to LSTM(BOC), thereby underscoring the crucial importance of modeling road network topology. However, its performance fluctuates within the complex Manhattan road network, achieving an EDS of 2.9163 km and revealing limitations in modeling highly intricate spatial correlations.

Under these circumstances, the proposed H-STQGCN algorithm exhibits both outstanding and stable performance. It achieves an optimal EDS of 2.0423 km on the Porto dataset, representing an improvement of approximately 0.9% over ST-GCN. For the San Francisco dataset, its EDS of 2.1414 is comparable to that of ST-GCN (2.1573). For the New York dataset, the EDS reaches 2.6282 km, representing a substantial 9.9% reduction in error relative to ST-GCN and an approximately 3.3% improvement over the next-best LSTM(BOC).

This consistent performance advantage demonstrates that, by deeply integrating the quantum computing potential validated by QLSTM with the robustness of classical spatio-temporal graph convolutions, the proposed algorithm can more effectively capture high-dimensional spatial correlations among nodes, thereby maintaining superior predictiveaccuracy and robustness in complex urban environments.

b) RMSE experimental results

Fig 7 Comparison of EDS and RMSE among the three cities for each algorithm. The results further present high consistency with two metrics, which also support our previous conclusions.

As illustrated in TABLE III, H-STQGCN obtains the RMSE of 2.3134 km, 2.5915 km, and 3.1608 km for Porto, San Francisco, and Manhattan, New York, respectively. It is evident that H-STQGCN shows significant improvements over traditional methods and classical spatio-temporal convolutional networks. Particularly, for Manhattan, New York, H-STQGCN obtains an RMSE significantly smaller than that of ST-GCN, which is relative improvement reaches 14.1%. H-STQGCN also shows improvement over QLSTM, which demonstrates that the hybrid algorithm can not only absorb the quantum advantages, but also overcome the spatial modeling limitations of purely quantum temporal algorithms.

It offers strong empirical evidence of the stability and superiority of the algorithm in another critical scenario.

*c) Loss Analysis*

Fig 8 Curve of cross-entropy loss during the training process of the algorithm.It can be seen from Fig. 8 that when the training period increases, both the training loss and validation loss will decrease rapidly, and gradually approach a certain value after about ten periods. In addition, the two curves do not diverge significantly, which means that the algorithm not only can learn the information of data, but also can reduce the serious overfitting. The above results verify that the proposed algorithm structure and training strategy are effective and stable, and the generalization ability of the algorithm is strong.

*d) Ablation Experiment*

To further quantify the contributions of the BOC semantic vector and the quantum branch QGCN—the core components of H-STQGCN—ablation experiments were conducted within the same framework by removing or substituting specific modules. The results are summarized in TABLE IV.

An analysis of TABLE IV indicates that the core components and synergistic interactions of H-STQGCN substantially enhance predictive performance while also demonstrating robust adaptability in complex scenarios. The effectiveness of the BOC semantic vector is thereby fully validated. By comparing the experiments using GCN+TCN (without BOC) and GCN+TCN (BOC), it is observed that the introduction of this feature enhances algorithm performance across all three cities. In San Francisco, the EDS decreased by 6.4%, while in Manhattan, New York, it decreased by 3.3%. This finding confirms that incorporating POI functional attribute information enriches the semantic context and deepens the understanding of urban functional zoning, thereby enhancing predictive accuracy.

The breakthrough contribution of the quantum graph convolutional module is particularly noteworthy. When comparing the results of GCN+TCN (with BOC) and GCN+QGCN+TCN (with BOC), the RMSE decreased by 3.2% in Porto, whereas Manhattan, New York, exhibited a 9.9% reduction in EDS and a 14.1% improvement in RMSE. This result fully validates the capability of quantum computing to efficiently explore high-dimensional feature spaces and accurately capture complex non-local spatial correlations. The synergistic enhancement effect among components is shown to be crucial.

The complete algorithm obtains the best performance at Porto. The EDS and RMSE of complete algorithm are 2.0423 km and 2.3134 km, respectively, which are improved by 1.3% and 3.7% compared with baseline algorithm (GCN+TCN without BOC). It demonstrates a synergistic effectiveness. The complete algorithm is applied on Manhattan, New York with complex road network. The EDS and RMSE are improved by 9.9% and 14.1% respectively. It also demonstrates the superiority of this hybrid algorithm on highly complex spatio-temporal correlation problem.

## V. CONCLUSION

To address the challenge of capturing global spatial dependencies in complex road networks for accurate taxi destination prediction, we propose the H-STQGCN algorithm. By integrating PQC-enhanced quantum layers for high-dimensional feature mapping with classical GCN and temporal convolution modules, our framework effectively unifies non-local spatial extraction with dynamic temporal modeling. Extensive experiments demonstrate that H-STQGCN significantly outperforms mainstream benchmarks, including MMLP, LSTM, ST-GCN, and QLSTM. It is noteworthy that our model demonstrates a 14.05% reduction in RMSE and a 9.88% reduction in EDS compared to ST-GCN within the complex Manhattan road network. Furthermore, it achieves the lowest recorded error rates on the Porto dataset (EDS: 2.0423, RMSE: 2.3134). Our results confirm that our approach substantially mitigates limitations of classical methods when applied to complex urban environments.

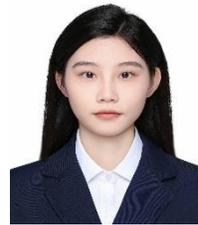

**Xiuying Zhang** received the B.S. degree in 2024. She is currently pursuing the master's degree with the School of Physics, University of Electronic Science and Technology of China. Her research interests include quantum computing, intelligent transportation, and quantum machine learning.

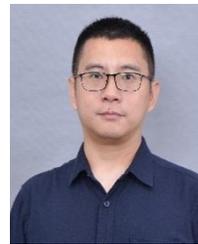

**Qinsheng Zhu** received the Ph.D. degree in Science from Sichuan University, Chengdu, China, in 2008. He is currently an Associate Professor with the School of Physics, University of Electronic Science and Technology of China, Chengdu, China. His research interests include quantum information, quantum computing, quantum machine learning, edge computing, and the Internet of Things..

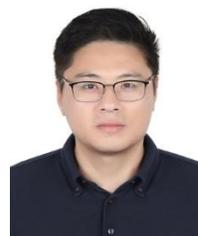

**Xiaodong Xing** received the Ph.D. degree from Paris-Saclay University, France. He is currently an assistant professor in the School of Quantum Information Future Technology, Henan University, China. His research interests include the classical-quantum algorithms, ultracold molecules and ultracold atom-molecule collisions.